\documentclass[aip,jcp,twocolumn,reprint,longbibliography]{revtex4-1}
\usepackage{epsfig}
\usepackage{color}
\usepackage{amsmath}
\usepackage{amsfonts}
\usepackage{natbib}
\usepackage{notes2bib}
\usepackage{chemformula}
\usepackage{verbatim}
\usepackage{upgreek}
\usepackage{multirow}
\usepackage{hhline}

\newcommand{\degree}{^{\circ} }

\newcommand{\avg}[1]{\left<#1\right>}

\newcommand{\brac}[1]{\left[#1\right]}

\newcommand{\para}[1]{\left(#1\right)}

\newcommand{\etal}{\emph{et al.}}

\newcommand{\kT}{\ensuremath{k_{\rm B}T}}

\newcommand{\nb}{\ensuremath{\mathbf{n}}}

\newcommand{\vb}{\ensuremath{\mathbf{v}}}

\newcommand{\ace}{{C$_2$H$_2$}}
\newcommand{\amo}{{NH$_3$}}

\newcommand{\abint}{{ab initio}}
\newcommand{\chn}{{C-H$\cdots$N }} 
\newcommand{\nhpi}{{N-H$\cdots\pi$ }} 
\newcommand{\aacrystal}{{acetylene:ammonia (1:1)}}

\usepackage{color}

\begin{document}


\title{Molecular Structure, Dynamics, and Vibrational Spectroscopy of the Acetylene:Ammonia (1:1) Plastic Co-Crystal at Titan Conditions}


\author{Atul C. Thakur}
\author{Richard C. Remsing}
\email[]{rick.remsing@rutgers.edu}
\affiliation{Department of Chemistry and Chemical Biology, Rutgers University, Piscataway, NJ 08854}




\begin{abstract}
The Saturnian moon Titan has a thick, organic-rich atmosphere, and condensed phases of small organic molecules
are anticipated to be stable on its surface. 
Of particular importance are crystalline phases of organics, known as cryominerals, which can play
important roles in surface chemistry and geological processes on Titan. 
Many of these cryominerals could exhibit rich phase behavior,
especially multicomponent cryominerals whose component molecules have multiple solid phases.
One such cryomineral is the \aacrystal~co-crystal, and here we use density functional theory-based ab initio molecular dynamics simulations to quantify its structure and dynamics at Titan conditions. 
We show that the \aacrystal~co-crystal is a plastic co-crystal (or rotator phase) at Titan conditions because the ammonia molecules are orientationally disordered. 
Moreover, the ammonia molecules within this co-crystal rotate on picosecond timescales, and this rotation
is accompanied by the breakage and reformation of hydrogen bonds between the ammonia hydrogens and the $\pi$-system of acetylene. 
The robustness of our predictions is supported by comparing the predictions of two density functional approximations at different levels of theory, as well as through the prediction of infrared and Raman spectra that agree well with experimental measurements. 
We anticipate that these results will aid in understanding geochemistry on the surface of Titan. 
\end{abstract}

\maketitle

\raggedbottom

\section*{Introduction}

Titan, the largest moon of Saturn, is of great interest to planetary science and prebiotic chemistry communities because of its similarities to Earth~\cite{coustenis2008titan, raulin2012prebiotic, mackenzie2021titan, sagan1992titan, cable2012titan}. 
Like Earth, Titan has a thick atmosphere, but it is composed mainly of nitrogen and methane, and
atmospheric photochemistry creates a rich inventory of organic molecules on Titan~\cite{cable2012titan, horst2017titan, waite2007process, lavvas2008coupling1, lavvas2008coupling2}.
Titan's low surface temperature of approximately 94~K --- a major difference between Titan and Earth --- leads to condensation of organic molecules from the atmosphere and onto the surface. 
The surface of Titan has stable liquids composed mainly of hydrocarbons, which undergo seasonal rainfall cycles analogous to water-based cycles on Earth~\cite{hayes2016lakes}. 
In addition to organic liquids, organic solids can form through condensation or surficial processes at Titan surface temperatures.
Indeed, data from the Cassini-Huygens mission has provided strong evidence to suggest that Titan's surface is dominated by organic solids~\cite{lorenz2008titan, lorenz2006sand, elachi2005cassini, paganelli2007titan}.
These organic crystals, also referred to as cryominerals, could play a major role in geology, geochemistry, and even prebiotic chemistry on Titan's surface, similar to the importance of Earth's minerals in the terrestrial analogs of these processes~\cite{cable2021titan, maynard2018prospects, lunine2020astrobiology}.

Titan's cryominerals are an exciting class of crystalline materials that exist as pure phases or multicomponent solids composed
of more than one type of organic molecule --- co-crystals~\cite{cable2021titan,maynard2018prospects, maynard2016co}.
These molecular co-crystals are typically soft because they are held together mainly
by relatively weak intermolecular interactions, such as hydrogen bonding, $\pi$-$\pi$, or van der Waals interactions. 
These different types of intermolecular interactions act at different energy scales, such that molecular crystals often
display rich phase behavior consistent with thermal excitations disrupting various interactions at different temperatures. 
For example, molecular crystals are translationally and orientationally ordered at low temperatures, resulting in a crystal phase. 
However, as the temperature is increased, orientational degrees of freedom can become disordered without the solid melting. 
As a result, the molecular solid is translationally ordered but orientationally disordered, resulting in a `rotator' or `plastic crystal' phase~\cite{klein1990simulation}.
This complex phase behavior of cryominerals must be quantified to gain an understanding of their role in surface processes on Titan.
For example, a plastic crystal typically displays different mechanical, thermodynamic, and chemical properties than those of the perfect crystal. 
Crystal plasticity usually softens elastic constants and increases mechanical flexibility~\cite{lynden1994translation}.
Anomalies in the density, thermal expansivity, specific heat, and electrical properties of molecular solids have also been directly connected to the presence of plastic phases~\cite{maynard2018prospects, nose1983study}.  
The dynamic disorder within a plastic phase can facilitate transport through a crystal via molecular paddlewheel mechanisms~\cite{klein1990simulation,scholz2022superionic, alarco2004plastic, zhao2020designing, mondal2016proton, mondal2016thermal}. 
Seismic discontinuities across Earth's mantle have also been related to the plastic crystalline phase within deep Earth minerals~\cite{chaplot2001molecular}.   
Together, these effects may ultimately influence metamorphism, fracturing, crack propagation, and erosion rates on Titan.
Connecting these molecular level insights to the large-scale surface features constitutes an essential step in constructing a foundational understanding of the geological evolution and possible prebiotic chemistry of Titan.~\cite{maynard2018prospects,lynden1994translation}

In this work, we investigate a model Titan cryomineral, the \aacrystal~co-crystal.
The acetylene:ammonia (1:1) co-crystal is a prime candidate to display a plastic phase because both components exhibit plastic crystal phases~\cite{albert1972diffusion,van1978single,luo1986raman, fortes2003hydrogen, eckert1984structure}. 
Moreover, the component molecules that make up the \aacrystal~co-crystal have been confirmed in Titan's atmosphere and on the surface by instruments onboard Cassini~\cite{cable2018acetylene, singh2016acetylene,niemann2010composition,waite2005ion,coustenis2007composition,cui2009analysis}.
Acetylene may be the second most abundant photochemical product in Titan's hazy atmosphere~\cite{czaplinski2020experimental,horst2017titan} and has been linked to evaporite deposits~\cite{cordier2009estimate,cable2018acetylene} and equatorial dunes~\cite{abplanalp2019low,lorenz2006sand,czaplinski2020experimental} on Titan's surface.  
Along with its pure form, acetylene is also known to co-crystallize with many small organic molecules forming a variety of potential Titan cryominerals~\cite{kirchner2010co,maynard2018prospects,cable2021titan}.  
Like acetylene, ammonia has been detected in Titan's atmosphere, although measurement difficulties have sparked a debate on these results~\cite{cui2009analysis,magee2009inms,cable2018acetylene}.
Titan's subsurface ocean is predicted to be a mixture of water and ammonia, which can erupt from the surface via cryovolcanism, outgassing, and geysering events that move ammonia from Titan's interior to its surface~\cite{gilliam2016formation,sohl2014structural,cable2018acetylene,choukroun2010thermodynamic,tobie2005titan}.
The co-existence of acetylene with ammonia, either in Titan's atmosphere or via deposition of ammonia-rich slurry on top of solid acetylene rich deposits, can quickly lead to the formation of a co-crystal that is stable under Titan's surface conditions and to methane/ethane fluvial and pluvial events~\cite{cable2018acetylene}.
As a consequence, the \aacrystal~co-crystal is likely to be found near geochemically and biologically interesting sites such as cryovolcanic cones, ammonia eruption pits, and surface flows, in addition to Titan's stratosphere.\cite{cable2018acetylene,cable2021titan, lopes2013cryovolcanism,nelson2009saturn} 
Here, we explore the structure and dynamics of the \aacrystal~co-crystal using ab initio density functional theory-based molecular dynamics simulations at  $T = 30$ K and $T = 90$ K, where the temperatures correspond to a crystal and plastic crystal, respectively.  
To identify static signatures of the plastic crystal phase, we quantify the translational and orientational structure
of the co-crystal and identify the orientational disorder in the plastic crystal phase. 
Through an examination of translational and rotational dynamics, we show that the \aacrystal~co-crystal exhibits rapid dynamic orientational disorder in the plastic phase. 
We connect the observed orientational disorder to the dynamics of \nhpi~type hydrogen bonding within the co-crystal.
We end with a discussion of the vibrational fingerprints of co-crystal formation while also reflecting on the accuracy of our results through comparison of experimentally-measured and computed vibrational spectra.  
We conclude by highlighting the commonality of plastic phases in Titan's potential cryominerals and put our results within the broader context of developing a thorough understanding of Titan's minerals.

\section*{Simulation Details}

We performed Born-Oppenheimer molecular dynamics (BOMD) simulations using density functional theory (DFT) within the QUICKSTEP~\cite{vandevondele2005quickstep} electronic structure module of the CP2K code~\cite{kuhne2020cp2k, hutter2014cp2k, vandevondele2003efficient}.
QUICKSTEP employs a dual atom-centered Gaussian and plane wave (GPW)~\cite{lippert1997hybrid,vandevondele2005quickstep} basis approach for representing wavefunctions and electron density, leading to an efficient and accurate implementation of DFT. 
We used the molecularly optimized (MOLOPT) Goedecker-Teter-Hutter (GTH) triple-$ \zeta $ single polarization (TZVP-MOLOPT-GTH) Gaussian basis set~\cite{vandevondele2007gaussian} for expanding orbital functions, along with a plane wave basis set with a cutoff of 500~Ry for representing the electron density. 
The core electrons were represented using GTH pseudopotentials~\cite{krack2005pseudopotentials,hartwigsen1998relativistic,goedecker1996separable}.  
Exchange-correlation (XC) interactions were approximated using the Perdew-Burke-Ernzerhof (PBE) generalized gradient approximation (GGA) to the exchange-correlation functional~\cite{perdew1996generalized}, as implemented in CP2K. 
To account for long-ranged dispersion interactions, we used Grimme's D3 van der Waals correction (PBE+D3)~\cite{grimme2010consistent,grimme2011effect}.
To assess the accuracy of the density functional approximation, we also performed simulations using the r$^2$SCAN meta-GGA~\cite{furness2020accurate} combined with the appropriate non-local rVV10 correction for the long-range dispersion effects (r$^2$SCAN+rVV10)~\cite{sabatini2013nonlocal, ning2022workhorse}.
We determined a planewave cutoff of 700 Ry to be sufficient for describing the system with r$^2$SCAN+rVV10.   
We modeled a $2\times2\times2$ supercell of the \aacrystal \ co-crystal using the structure reported by Boese \etal~(Cambridge Structural Database ID: FOZHOS)~\cite{boese2009synthesis}.
We equilibrated the system for at least 10~ps in the canonical (NVT) ensemble at constant temperatures of 30~K and 90~K using the canonical velocity rescaling (CSVR) thermostat~\cite{bussi2007canonical}.
For the calculation of dynamic properties, equations of motion were propagated in the microcanonical (NVE) ensemble for approximately 50~ps using a velocity Verlet integrator with a timestep of 1~fs. 
Maximally localized Wannier functions (MLWFs) were obtained on-the-fly, minimizing the spreads of the MLWFs according to the general and efficient formulation implemented in CP2K~\cite{marzari2012maximally, berghold2000general, marzari1997maximally}.
The centers of the computed MLWFs were used to define molecular dipoles, and polarizability tensors which were then used to extract IR and Raman signatures from the generated BOMD trajectories using the TRAVIS analyzer~\cite{thomas2013computing,brehm2012travis,brehm2020travis}.
The IR and the Raman spectra were computed from a 20~ps long trajectory with MLWFs obtained at every step. 
The matrix elements of the full Raman polarizability tensor were extracted from the displacements of the MLWF centers recomputed by applying an electric field of strength 0.0005 \emph{a.u.} along the crystal $x$, $y$, and $z$ directions respectively. 
Each simulation with an electric field also resulted in production runs of 20~ps, such that the Raman spectrum is estimated from a total of 80~ps of simulation (four 20~ps trajectories). 
The IR spectra were also computed for pure solid acetylene and solid ammonia.
Solid acetylene undergoes a phase transition into an orthorhombic phase at 133~K. 
However, to the best of our knowledge, the crystal structure of orthorhombic solid acetylene has not been determined. 
Instead, we used the crystal structure for the orthorhombic phase of dideuteroacetylene (C$_2$D$_2$) (Cambridge Structural Database ID: ACETYL07) as a close approximation, and we replaced the deuterium atoms with hydrogens in our simulations~\cite{koski1975neutron}. 
For solid ammonia, we similarly used the crystal structure for cubic trideuteroammonia (ND$_3$) (Cambridge Structural Database ID: 34244)~\cite{reed1961neutron}.

\section*{Results and Discussions}

\subsection*{Translational Order and Orientational Disorder at Titan Conditions}

\begin{figure}[tb]
\begin{center}
\includegraphics[width=0.48\textwidth]{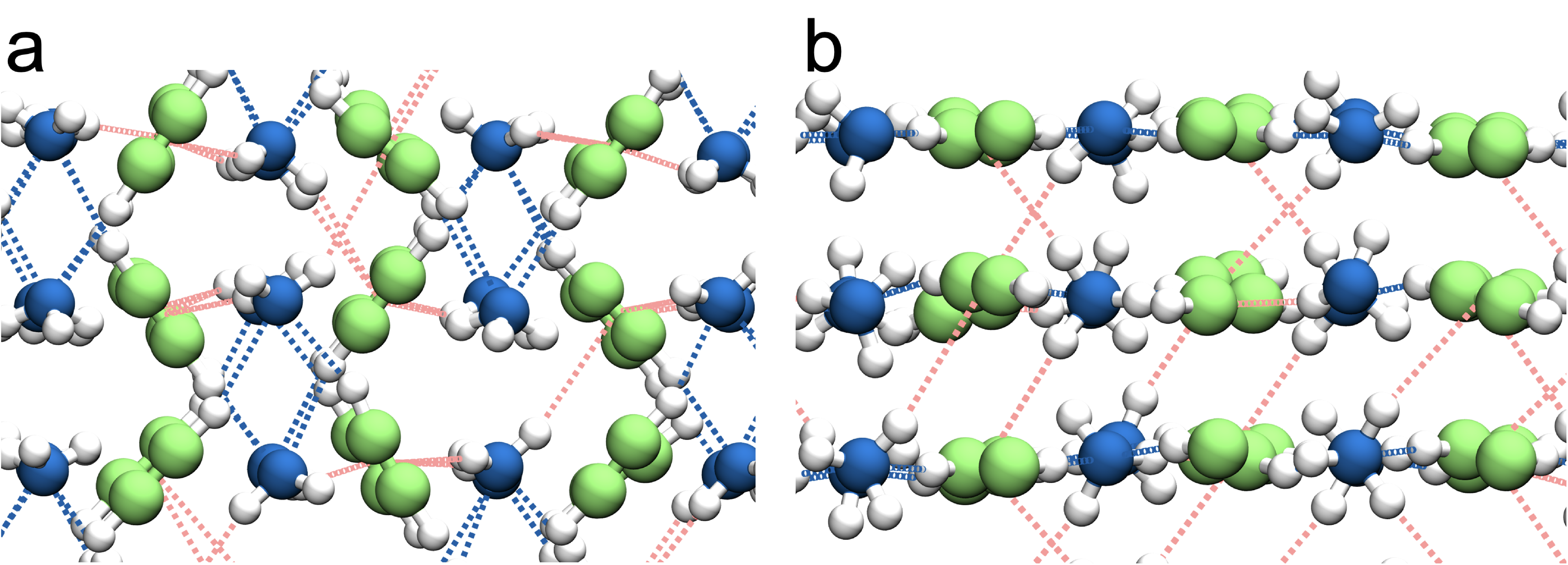}
\end{center}
\caption
{
Snapshot illustrating the zig-zag geometry of the \aacrystal~co-crystal with \chn~and \nhpi~type hydrogen bonding interactions shown as dotted lines. 
The hydrogen bonds are colored according to the donor atom.
Views from both (a) crystal $z$ and (b) crystal $y$ axes are shown.
}
\label{fig:aasnap}
\end{figure}

The \aacrystal~co-crystal has a layered structure consisting of antiparallel planes of zigzag chains,
as shown in Figure~\ref{fig:aasnap}~\cite{cable2018acetylene,preston2012formation,boese2009synthesis, cable2021titan}. 
The layered arrangement within the co-crystal is primarily held together via a network of hydrogen bonds and weak van der Waals interactions. 
We find two types of hydrogen bonds in the \aacrystal~co-crystal: \chn~and \nhpi~hydrogen bonds~\cite{cable2018acetylene,preston2012formation,boese2009synthesis},
shown as blue and red dashed cylinders in Fig.~\ref{fig:aasnap}, respectively.
The \chn~hydrogen bonds are formed by electrostatic attractions between the lone pair electrons of the nitrogen atom and the H atoms of acetylene molecules.  
Similarly, electrostatic attractions between the H atoms of ammonia and the electron-rich $\pi$-system of a neighboring acetylene molecule give rise to \nhpi~hydrogen bonds. 
These \chn~and \nhpi~interactions play a crucial role in determining the structure and dynamics of the \aacrystal~co-crystal.
%

\begin{figure}[tb]
\begin{center}
\includegraphics[width=0.48\textwidth]{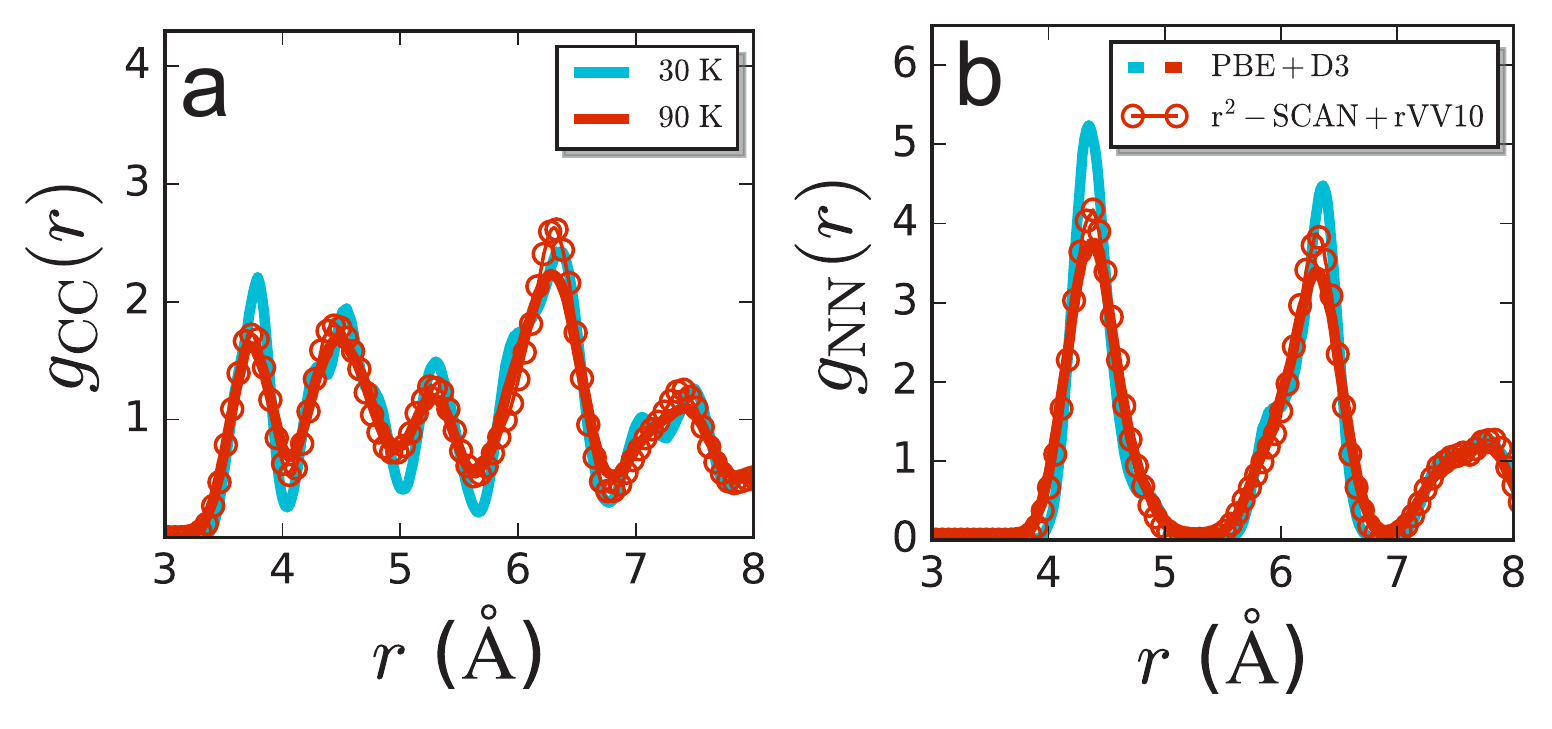}
\end{center}
\caption
{
Site-site radial distribution functions (RDFs), $g(r)$, computed for (a) acetylene carbons and (b) ammonia nitrogens within the \aacrystal~co-crystal at $T=30$~K and $T=90$~K. 
The solid lines indicate the RDFs evaluated with PBE+D3, and the circles correspond to those evaluated with r$^{2}$-SCAN+rVV10.  
}
\label{fig:plotrdf}
\end{figure}

\begin{figure}[tb]
\begin{center}
\includegraphics[width=0.48\textwidth]{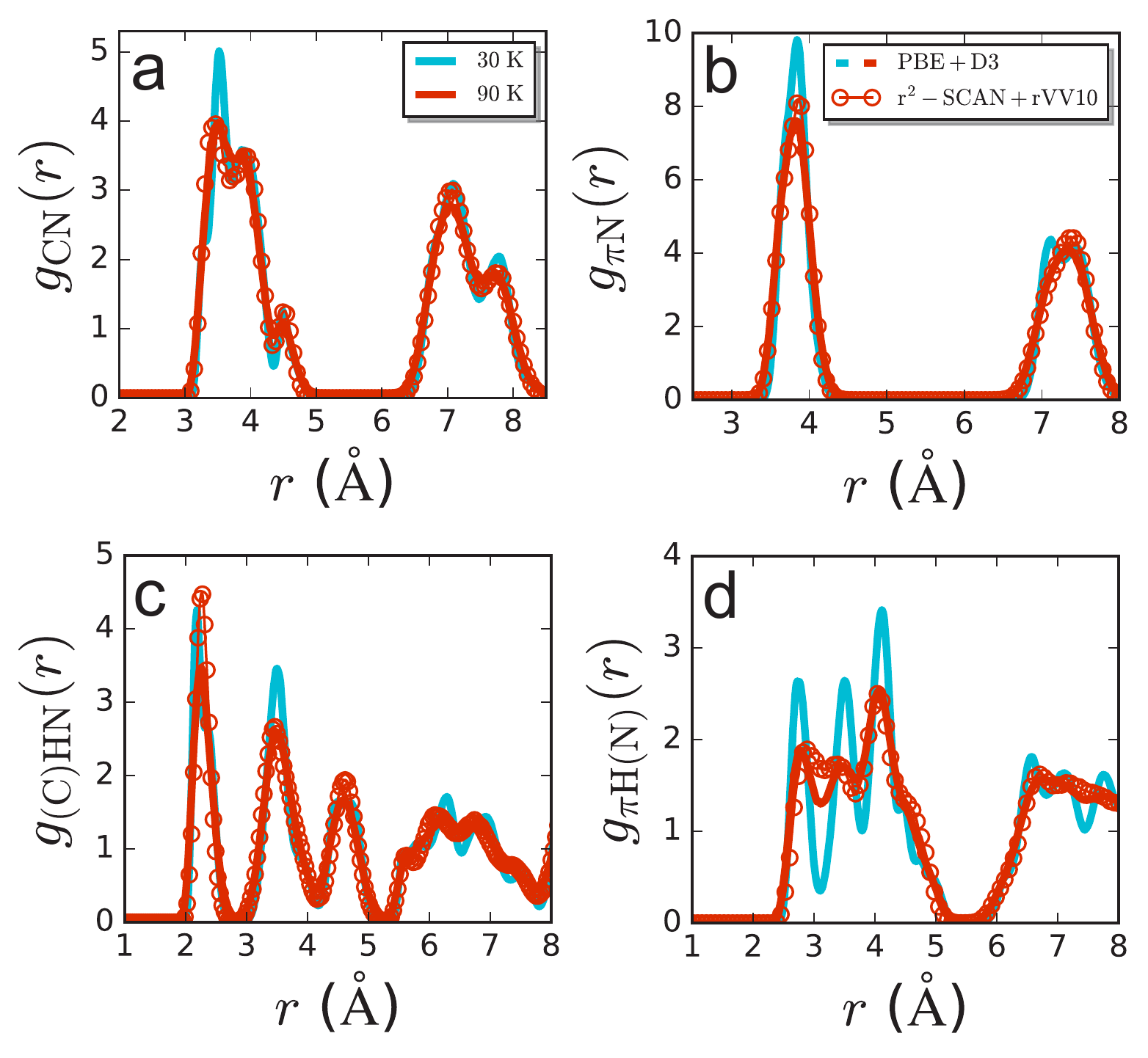}
\end{center}
\caption
{
Site-site radial distribution functions (RDFs), $g(r)$, for
(a) C-N, (b) $\pi$-N, (c) (C)H-N, and (d) $\pi$-H(N) correlations involved in (a,c) \chn~and (b,d) \nhpi~hydrogen bonds
at temperatures of 30~K and 90~K.
The solid lines indicate the RDFs evaluated with PBE+D3, and the circles correspond to those evaluated with r$^{2}$-SCAN+rVV10.  
}
\label{fig:hbrdf}
\end{figure}

%
We quantify the structure of the co-crystal by computing site-site radial distribution functions (RDFs), $g_{XY}(r)$~\cite{allen2017computer,chandler1987introduction},
where $X$ and $Y$ refer to atomic sites or the midpoint of the C-C bond when representing the acetylene $\pi$ electrons.
The periodic arrangement of carbon and nitrogen atoms within the co-crystal is reflected in the periodic nature of C-C and N-N RDFs (Figure~\ref{fig:plotrdf}). 
Integration of $g_{\rm CC}(r)$ and $g_{NN}(r)$ over the first peak yields coordination numbers of approximately 4 and 6, respectively,
consistent the arrangement of atoms within the co-crystal geometry. 
Increasing temperature from 30~K to 90~K results in broader, lower intensity peaks in the RDFs, consistent with increased  motion in the crystal due to thermal excitation. 
These results are largely independent of density functional approximation, although the r$^2$-SCAN+rVV10 functional gives slightly larger and narrower peaks in the C-C and N-N RDFs than those predicted by PBE+D3.
The hydrogen bonds between acetylene and ammonia molecules are reflected in the various RDFs quantifying correlations
between hydrogen bonding groups, Fig.~\ref{fig:hbrdf}.
The \chn~hydrogen bonds consist of acetylene donating a hydrogen bond to ammonia. 
Therefore, these hydrogen bonds can be quantified through C-N and (C)H-N RDFs, where (C)H is used to indicate that the hydrogen is bonded to a carbon. 
The C-N RDF, $g_{\rm CN}(r)$, exhibits a large first peak near 3.5~\AA \ (Figure~\ref{fig:hbrdf}a), the integral of which gives a coordination number of two.
This peak corresponds to carbon atoms of two different acetylene molecules that are H-bonding with ammonia, highlighted by the snapshots in Figure~\ref{fig:aasnap}.
There is a second and third peak soon after (before 5~\AA), both of which correspond to carbons that are not involved in H-bonds with the central ammonia molecule.
This structure is further emphasized by the (C)H-N RDFs, which exhibit a peak at approximately 2.25~\AA, suggesting that acetylene hydrogens point directly at the lone pair of the nitrogen to form \chn~hydrogen bonds (Fig.~\ref{fig:hbrdf}c). 
Integration of the first peak in this RDF yields a coordination number of two, further suggesting that the nitrogen atom is, on average, engaged in hydrogen bonding with two acetylene molecules. 
To characterize \nhpi~hydrogen bonds, we need a site to represent the $\pi$ system of each acetylene molecule. 
Here, we represent the location of the $\pi$ system as a single site located at the midpoint of the C-C triple bond, but note that
similar results are found if the centers of maximally localized Wannier functions are used to represent the location of $\pi$ electrons.
The $\pi$-N RDF displays a single, well-defined peak just before 4~\AA,
suggesting a single H-bond donated from an ammonia molecule to an acetylene molecule, Fig.~\ref{fig:hbrdf}b,
as shown in the snapshots in Fig.~\ref{fig:aasnap}.
The $\pi$-H(N) RDF displays three peaks below 4.5~\AA, corresponding to the three hydrogens of the ammonia molecule, Fig.~\ref{fig:hbrdf}d. 
The first peak located at distances smaller than 3~\AA \ results from H atoms involved in a hydrogen bond with acetylene. 
Increasing temperature from 30~K to 90~K weakens hydrogen bonds, as evidenced by the broader and less intense peaks in the RDFs.
Importantly, heating the \aacrystal~co-crystal to 90~K significantly increases the disorder observed in the $\pi$-H(N) RDF, Fig.~\ref{fig:hbrdf}d.
The deep minima between peaks, especially that between the first and second peak, increase to be comparable in height to some of the peaks. 
This indicates that ammonia H atoms have a significantly higher probability of occupying these interstitial regions, suggesting that the ammonia molecules are orientationally disordered at 90~K.
We examine the dependence of the H-bond structure on density functional approximation at 90~K.
We find that the first peak in the (C)H-N RDF is significantly larger in the r$^2$-SCAN+rVV10 system than in the PBE+D3 one, such that the peak height is larger than that at 30~K (with PBE+D3). 
This suggests that \chn~hydrogen bonds are stronger in the r$^2$-SCAN+rVV10 system. 
In contrast, the $\pi$-H(N) RDF is more disordered in the r$^2$-SCAN+rVV10 system, suggesting that \nhpi~hydrogen bonds are weaker and the ammonia molecules are more disordered than in the PBE+D3 system.
The weaker \nhpi~hydrogen bonds produced by the r$^2$-SCAN+rVV10 functional may be due to increased charge transfer along the \chn~interaction
suggested by the stronger \chn~hydrogen bonds. 
%

\begin{figure*}[tb]
\begin{center}
\includegraphics[width=0.98\textwidth]{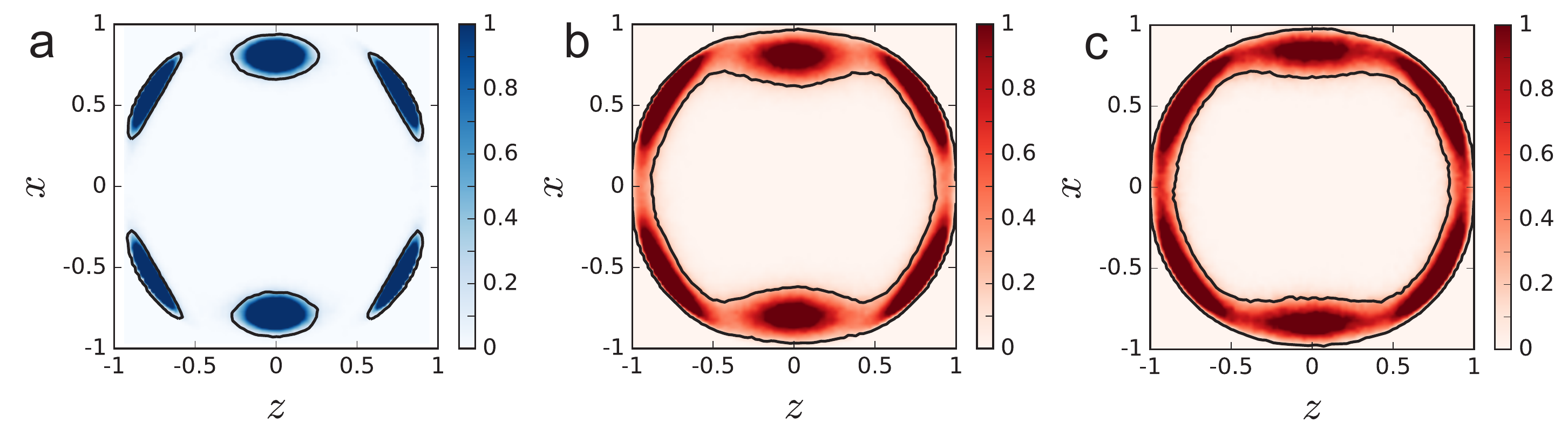}
\end{center}
\caption
{
Probability distributions of the ammonia H atom positions projected onto the $ac$ plane
at (a) $T=30$~K and (b,c) $T=90$~K. 
Results for both (b) PBE+D3 and (c) r$^2$SCAN+rVV10 functionals are shown at 90~K. 
The contour line is drawn at a probability of 0.2. 
}
\label{fig:plotos}
\end{figure*}

%
The RDFs discussed above are averaged over orientational degrees of freedom and, therefore, cannot directly quantify orientational order in the \aacrystal~co-crystal. 
To quantify the orientational structure of ammonia molecules within the co-crystal, we compute the probability distribution of  ammonia H atom positions projected on the $xz$ plane of the crystal~\cite{welchman2014positional}.
At 30~K, the ammonia molecules are orientationally ordered, as reflected by the distinct, disconnected regions of high probability, Fig.~\ref{fig:plotos}a. 
The six distinct peaks in the probability distribution are consistent with the two types of configurations of ammonia molecules
shown in Fig.~\ref{fig:aasnap}a, which are orientationally ordered but point their dipole moments (and hydrogens) in opposite directions.
In contrast, the ammonia molecules show evidence of orientational disorder at 90~K; the probability distribution is continuous along an ellipse in the $xz$ plane, Fig.~\ref{fig:plotos}b.
The appearance of non-zero probability between the peaks in the distribution at 90~K indicates that ammonia molecules can adopt all orientations in the $xz$ plane, although the six peaks are still the most populated orientations.
Similar results are obtained at 90~K using the r$^2$-SCAN+rVV10 functional, although the distribution is broader
because the ammonia molecules are more disordered, Fig.~\ref{fig:plotos}c, as anticipated from the weaker \nhpi~hydrogen bonds in this system.
Therefore, in the temperature range on Titan's surface, our results suggest that the \aacrystal~co-crystal is in an orientationally disordered phase --- the \aacrystal~plastic co-crystal. 

\subsection*{Translational and Orientational Dynamics}

\begin{figure}[tb]
\begin{center}
\includegraphics[width=0.48\textwidth]{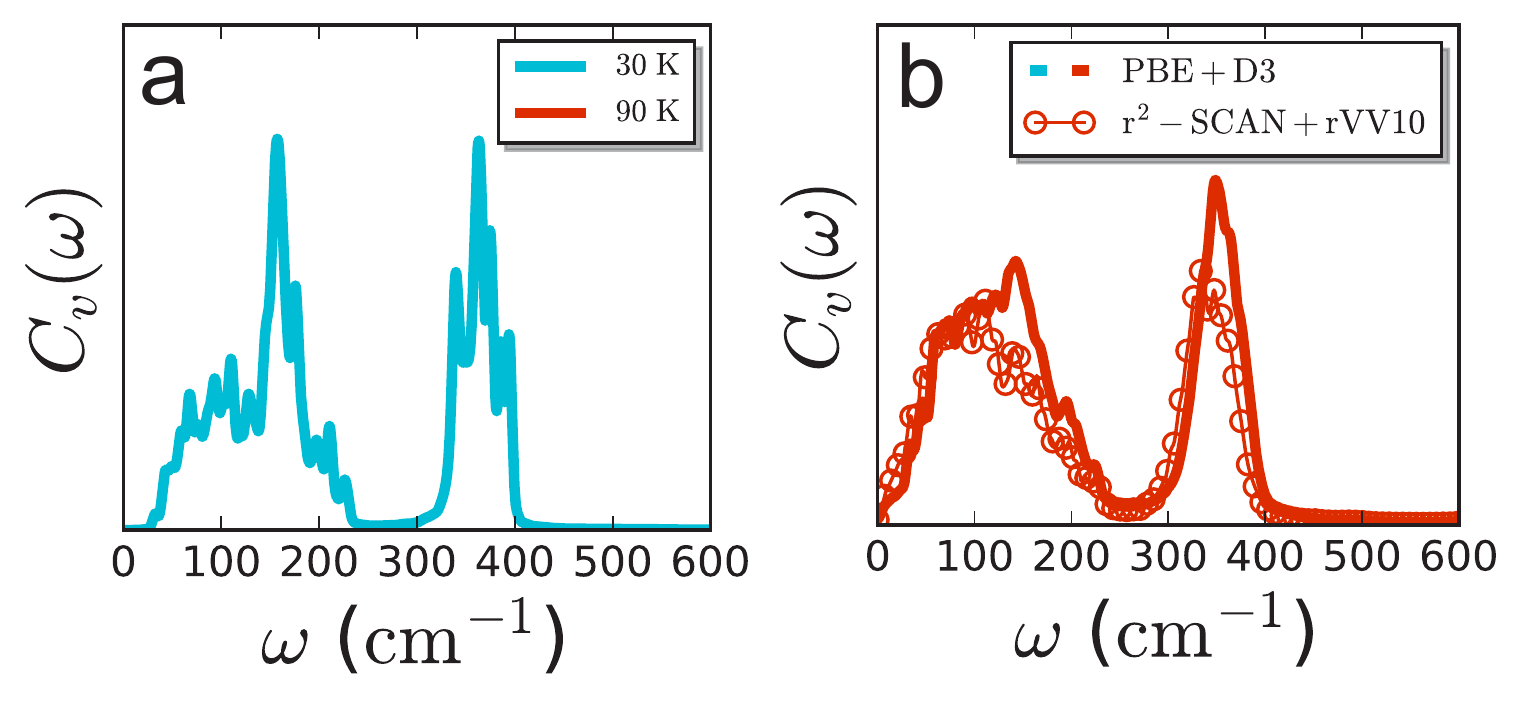}
\end{center}
\caption
{
Low frequency region of the total all-atom vibrational power spectrum, $\tilde{C}_v(\omega)$, evaluated for \aacrystal~co-crystal at (a) $T=30$~K and (b) $T=90$~K. 
Panel (b) also compares the vibrational power spectrum at $T=90$~K computed using two different density functional approximations.  
}
\label{fig:Plotvacf}
\end{figure}

The structural analysis above gives no information about molecular dynamics within the co-crystal,
and therefore cannot suggest whether or not the orientational disorder is dynamic. 
Therefore, we also quantify the translational and orientational dynamics of molecules in the \aacrystal~co-crystal.
We first quantify the translational dynamics of the co-crystal by computing normalized velocity auto-correlation functions,
\begin{equation}
C_v(t) = \frac{\avg{\sum_{i=1}^N\vb_i(t)\cdot \vb_i(0)}}{\avg{\sum_{i=1}^N\vb_i(0)\cdot \vb_i(0)}},
\label{VACF}
\end{equation}
where $\vb_i(t)$ is the velocity of atom $i$ at time $t$
and $\left\langle\cdots\right\rangle$ denotes an ensemble average. 
The power spectrum, ${C}_v(\omega)$, is obtained by Fourier transforming $C_v(t)$,
and corresponds to the vibrational density of states of the system, Fig.~\ref{fig:Plotvacf}.
The power spectrum at 90~K displays broader peaks than that computed at 30~K, reflecting the thermal excitation of more vibrational modes and increased disorder at higher temperatures.
The power spectra for both functionals used here show similar behavior, with slight differences in peak shapes.
The $\omega\rightarrow0$ limit of ${C}_v(\omega)$ is proportional to the diffusion coefficient~\cite{berne2009correlation}. 
Importantly, all three power spectra vanish in this limit,
indicating a lack of translational diffusion within the crystal, as expected for translationally ordered crystals. 
%

\begin{figure}[tb]
\begin{center}
\includegraphics[width=0.48\textwidth]{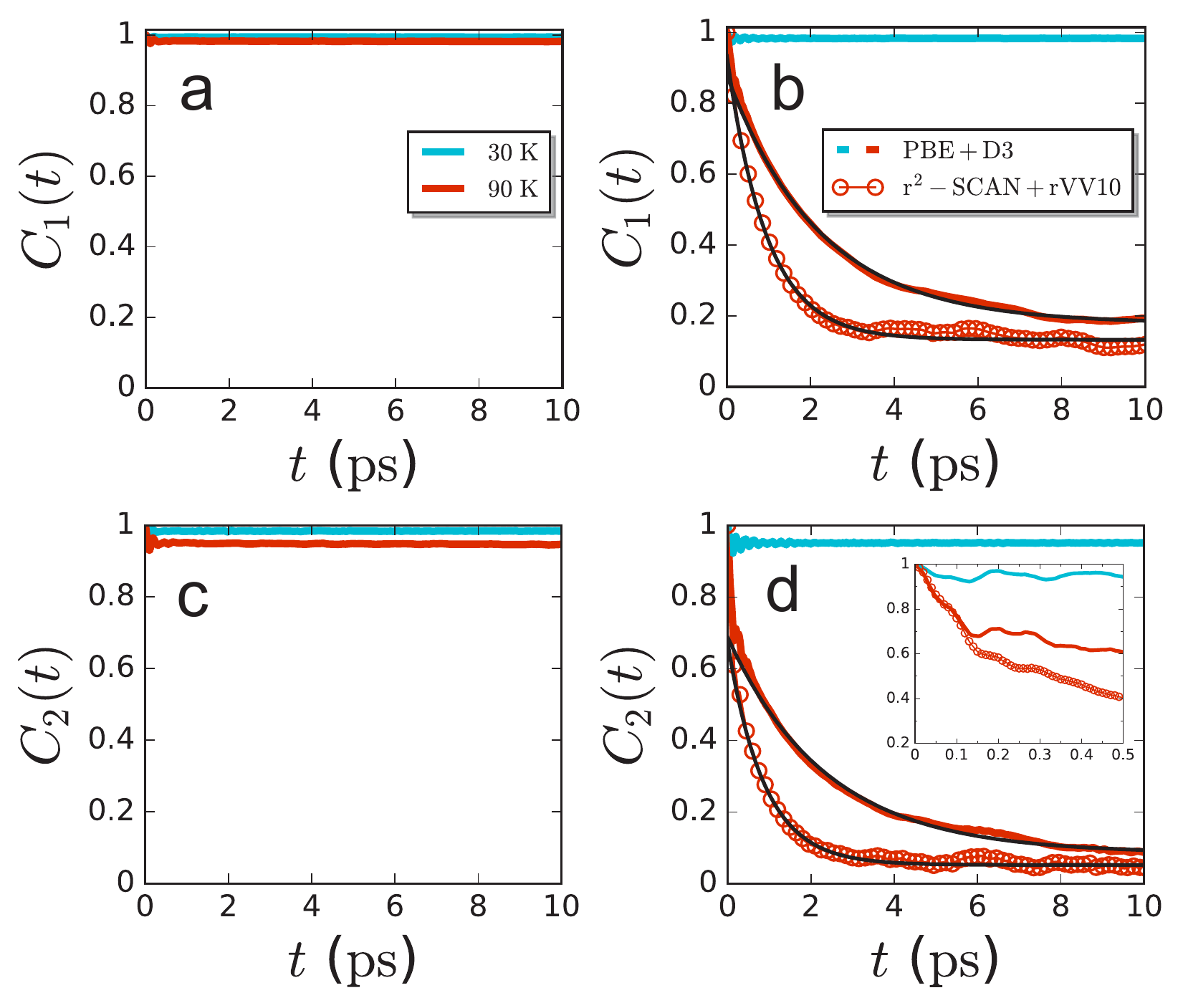}
\end{center}
\caption
{
Rotational time correlation functions, $C(t)$ and $C_2(t)$, for (a, c) \ace~and (b, d) \amo~molecules within the \aacrystal~co-crystal at $T=30$~K (cyan) and $T=90$~K (red).
The acetylene correlation functions are evaluated using the C-C bond vector, while those for ammonia molecules are based on the orientations of N-H bonds. 
Black solid lines indicate exponential fits to the correlation functions at long times. 
The inset in d shows the behavior of $C_2(t)$ from \amo~molecules at short times. 
}
\label{fig:C2RCF}
\end{figure}

%
We probe the orientational dynamics within the co-crystal by computing the rotational time correlation functions
\begin{equation}
C_n(t) = \avg{P_n(\nb(t)\cdot \nb(0))},
\end{equation} 
where $\nb(t)$ is an N-H bond vector for ammonia and a C-C bond vector for acetylene molecules at time $t$
and $P_n(x)$ is the $n$th order Legendre polynomial.
We focus on $C_1(t)$ and $C_2(t)$, which can be determined through spectroscopy~\cite{gordon1965molecular,gordon1965relations}.
Correlations in the orientational structure of acetylene molecules persist throughout the duration of the simulations at both 30~K and 90~K, Fig.~\ref{fig:C2RCF}a,c, suggesting that the acetylene molecules within the co-crystal are in a static crystalline phase and do not undergo any significant orientational dynamics.
Similar long-time correlations occur for ammonia molecules at 30~K, Fig.~\ref{fig:C2RCF}b,d, indicating an absence of any rotational dynamics within the co-crystal at low temperatures.
In contrast, $C_1(t)$ and $C_2(t)$ for ammonia molecules decay on picosecond timescales at 90~K.
This indicates that ammonia molecules exhibit dynamic orientational disorder.
We estimate the rotational correlation times $\tau_1$ and $\tau_2$ as the integral of the corresponding time correlation functions, $C_1(t)$ and $C_2(t)$, respectively. 
In practice, we split the correlation function into two parts at $t=\tau_{\rm cut}$ to perform the integration:
(i) the short time part from $t=0$ to $t=\tau_{\rm cut}$, which we numerically integrate (shown for $C_2(t)$ in the inset of Fig.~\ref{fig:C2RCF}d),
and
(ii) the rest of the correlation function, which we fit to an exponential function, $A_0 e^{-\tau_{\rm fit}/t}+A_1$.
Then, the rotational correlation time is given by
\begin{equation}
\tau_n = \int_0^{\tau_{\rm cut}} dt \brac{C_n(t)-C_n(\infty)} + A_0 \tau_{\rm fit} e^{-\tau_{\rm fit}/\tau_{\rm cut}}
\end{equation}
Using $\tau_{\rm cut}=0.5$~ps, we obtain rotational correlation times of $\tau_1=1.6\pm0.2$~ps for PBE+D3 and $\tau_1=0.77\pm0.01$~ps for r$^2$-SCAN+rVV10, and $\tau_2=1.5\pm0.1$~ps and $\tau_2=0.58\pm0.01$~ps for the PBE+D3 and r$^2$-SCAN+rVV10 systems, respectively.
This fast decay of ammonia rotational correlations suggests that the \aacrystal~co-crystal is in a dynamic orientationally disordered plastic phase at Titan surface conditions.  
%

%

%
Because the ammonia molecules rotate, but the acetylene do not, ammonia rotations are coupled to transient breakage and reformation of \nhpi~hydrogen bonds. 
To quantify the dynamics of \nhpi~hydrogen bonds, we compute the hydrogen bonding correlation function~\cite{luzar1996effect, luzar1996hydrogen, luzar2000resolving} 
\begin{equation}
 C_{\rm HB}(t) = \frac{\avg{ {h}(0){h}(t)}}{\avg{{h}}},
 \end{equation}
where the indicator function is given by
\begin{equation}
h(t) = \Theta\brac{ R^c - R(t)} \Theta\brac{ \phi^c - \phi(t) }
\end{equation}
and $\Theta(x)$ is the Heaviside function. 
With appropriate geometric definitions for a hydrogen bond, $h(t)=1$ if a hydrogen bond exists between a triplet of sites at time $t$ and $h(t)=0$ otherwise.
For \nhpi~hydrogen bonds, $R(t)$ and $\phi(t)$ correspond to the N$\cdots\pi$ distance and the angle formed by the N-H and N$\cdots\pi$ vectors, respectively; the $\pi$-site is taken to be the midpoint of the C-C bond. 
To define a \nhpi~hydrogen bond, we use a radial distance cutoff of $R^c=4$~\AA \ and an angular cutoff of $\phi^c=30\degree$.
For comparison, we also compute the hydrogen bond correlation function for \chn~hydrogen bonds, where $R(t)$ corresponds to a C-N distance,
$\phi(t)$ is the angle made by the C-N and (C)H-N vectors, $R^c=4$~\AA, and $\phi^c=30\degree$.
%

\begin{figure}[tb]
\begin{center}
\includegraphics[width=0.48\textwidth]{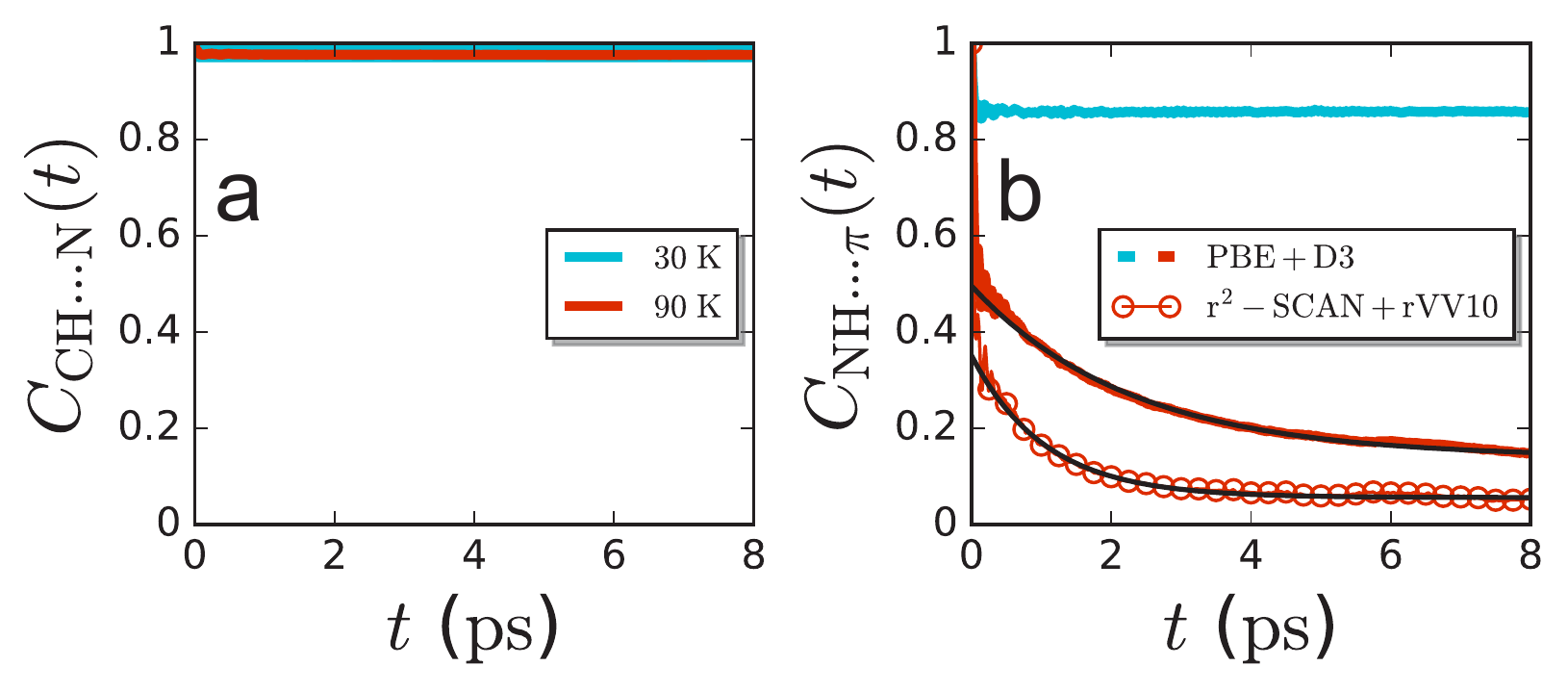}
\end{center}
\caption
{
Hydrogen bond auto-correlation functions, $ C_{HB}(t)$, evaluated for \chn as well as \nhpi type hydrogen bonding within the \aacrystal~co-crystal at (a) $T=30$~K and (b) $T=90$~K. 
The solid lines are results using the PBE + D3 XC functional, while the filled circles indicate the results from r$^{2}$-SCAN + rVV10 XC approximation. 
The solid black lines indicate exponential fits to the correlation functions. 
}
\label{fig:plothb}
\end{figure}

%
At 30~K, $C_{\rm HB}(t)$ does not decay for either type of hydrogen bond, Fig.~\ref{fig:plothb}a,b. 
The lack of decay indicates that there is no significant hydrogen bond breakage within the co-crystal at low temperatures. 
At 90~K, $C_{\rm HB}(t)$ also does not significantly decay for \chn~hydrogen bonds, Fig.~\ref{fig:plothb}a, indicating that these interactions remain intact for long periods of time. 
In contrast, $C_{\rm HB}(t)$ for \nhpi~hydrogen bonds at 90~K decay rapidly, suggesting that hydrogen bond breakage accompanies the rotational motion of ammonia molecules. 
We determine the hydrogen bond correlations time, $\tau_{\rm HB}$, from $C_{HB}(t)$ following the same approach used for rotational times.  
We find correlation times for \nhpi~hydrogen bond relaxation of $\tau_{\rm HB}=0.9\pm 0.06$~ps from PBE+D3
and $\tau_{\rm HB}=0.35\pm 0.01$~ps from r$^2$-SCAN+rVV10, slightly faster than the corresponding rotational times. 
This must be the case if hydrogen bond breakage accompanies rotation.
Moreover, the faster hydrogen bond dynamics of r$^2$-SCAN+rVV10 are consistent with weaker \nhpi~hydrogen bonds in this system. 
These faster hydrogen bond dynamics are also consistent with the faster rotational dynamics produced by r$^2$-SCAN+rVV10, further suggesting that hydrogen bond breakage accompanies rotational dynamics. 

\subsection*{Theoretical Vibrational Spectroscopy}

\begin{figure*}[tb]
\begin{center}
\includegraphics[width=0.98\textwidth]{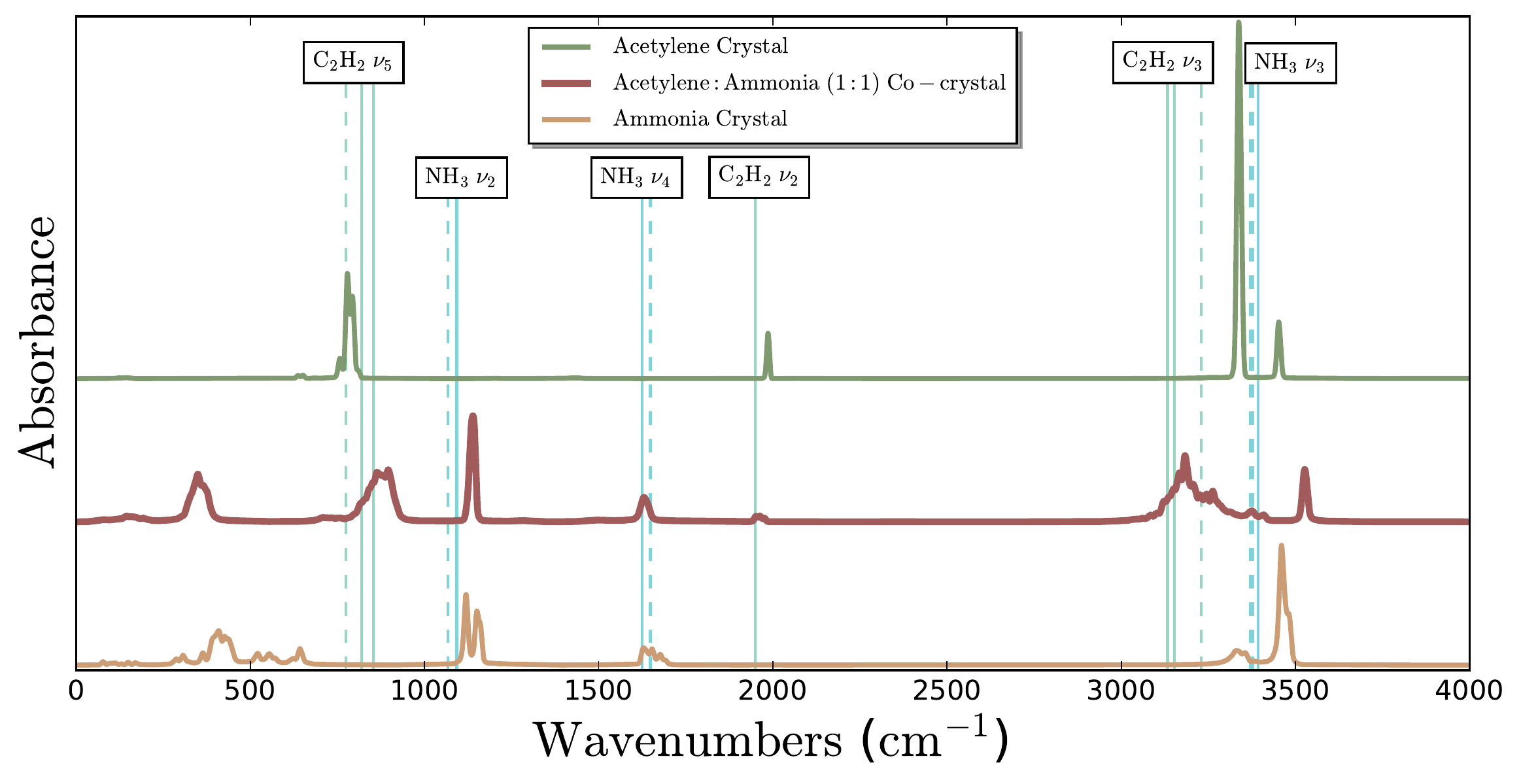}
\end{center}
\caption
{
Infrared (IR) spectra computed from simulations of orthorhombic solid acetylene, \aacrystal, and cubic solid ammonia crystals at $T=90$~K.
Vertical lines indicate experimentally measured peak positions~\cite{preston2012formation}; dotted lines correspond to pure acetylene and ammonia crystals, and solid lines correspond to co-crystal peaks.
}
\label{fig:IR}
\end{figure*}
%

\setlength{\tabcolsep}{8pt}
\begin{table*}[tb]
\centering
 \begin{tabular}{c  c  c  c  c c c } 
 \hline  
 & \multicolumn{6}{c}{Frequencies (cm$^{-1}$)} \\ 
 \hhline{~|-|-|-|-|-|-|} 
 &  \multicolumn{3}{c}{Calculated}  & \multicolumn{3}{c}{Experimental} \\ 
 \hhline{~---|||---}
 Band & Pure & Co-crystal & Shift & Pure & Co-crystal & Shift \\ [0.5ex] 
 \hline\hline
 \ace~$\nu_5$   & 778.7    & 863.2 & 84.5   &  774.4  & 819.1 & 44.7  \\  
                          &     -         & 895.8 & -   &              & 852.7  & -   \\
 \ace~$\nu_2$   & 1986.0  & 1962.6  & -23.4 &      -       & 1949.0 & -   \\
 \ace~$\nu_3$   & 3336.8  & 3166.8 & -170.0  & 3230.2  & 3132.9 & -97.3   \\
                          &     -         & 3183.1 & -  &       -       & 3152.7 & -  \\
 \amo~$\nu_2$  & 1119.7   & 1139.0 & 19.3 & 1066.6  & 1092.7 & 26.1   \\
 \amo~$\nu_4$  & 1655.9  & 1629.7 & -26.2  & 1648.1  & 1625.0 & -23.1   \\ 
 \amo~$\nu_3$  & 3460.0  & 3526.2 & 66.2 & 3369.5  & 3392.0 & 22.5   \\
                          & 3480.3  &     - & -         &  3377.3 &    - & -           \\
\hline
 \end{tabular}
 \caption{Calculated and experimental IR vibrational frequencies of \ace~and \amo~in pure orthorhombic solid acetylene, cubic solid ammonia, and \aacrystal~co-crystal, as well as the frequency shift induced by co-crystallization,
 given by the difference in the frequency of the co-crystal and the pure phases.
Experimental IR frequencies of pure and the co-crystalline phase are those reported by Preston~\etal~\cite{preston2012formation}.}
\label{tab:IR}
\end{table*}

%
Infrared (IR) and Raman spectroscopy are complimentary vibrational techniques that are well-suited for studying Titan cryominerals because of the sensitivity of molecular vibrations to their environment~\cite{cable2018acetylene, cable2020properties, vu2014formation, cable2019co, vu2022Buta, cable2021titan}. 
To provide molecular-scale interpretation of experimental spectra and to validate the accuracy of our models through comparison with experiments, we computed both the IR and Raman spectra of the \aacrystal~co-crystal. 
The IR absorption spectra were computed from the Fourier transform of the dipole derivative autocorrelation function defined according to~\cite{thomas2013computing},
\begin{equation}
A(\omega) \propto \int\avg{\dot{\mu}(0) \cdot \dot{\mu}(t)} e^{-i \omega t}~{\text{d}}t \label{eqIR}
\end{equation}
where $\mu(t) $ refers to the instantaneous dipole moment and $\dot{\mu}(t)$ is its time derivative.
The unpolarized Raman spectrum was computed as~\cite{thomas2013computing}
\begin{align}
I(\omega) &\propto \frac{(\omega_{\text{in}} - \omega)^4} {\omega} \frac{1}{ 1 -  e^{- \hbar \omega/\kT}} \nonumber \\
&\times \int 
\brac{\frac{ 45 a_p ^2(t) + 4\gamma_p^2(t)} {45}  + \frac{\gamma_p^2(t)} {15} }e^{-i \omega t}~{\text{d}}t \label{eqRaman}
\end{align}
where $ k_{\rm B} $ is Boltzmann's constant, $ \omega_{in} $ is the frequency of the incident light, $ \hbar $ is Plank's constant, and $ T $ is the temperature. 
The isotropic ($ a_p $) and anisotropic polarizabilities ($ \gamma_p $) are defined according to
\begin{equation}
a_p(t) =  \frac{1}{3}\para{ C_{xx}(t) + C_{yy}(t) + C_{zz}(t) }  \label{eqRaman3}
\end{equation}
and
\begin{equation}
\begin{split}
\gamma_p^2(t)  =  & \frac{1}{2}\para{C_{xx}(t) - C_{yy}(t)}^2 +  \frac{1}{2}\para{C_{yy}(t) - C_{zz}(t) }^2 \\ 
& +  \frac{1}{2}\para{ C_{zz}(t) - C_{xx}(t)}^2 + 3  \para{ C_{xy}(t) }^2 \\ 
& + 3 \left( C_{yz}(t) \right)^2 + 3 \left(  C_{zx}(t) \right)^2,
\end{split} \label{eqRaman4}
\end{equation}
where the $C_{ab}(t)$ indicates the time correlation function of the time derivatives of the polarizability, $\alpha_{ab}(t)$,
\begin{equation}
 C_{ab}(t) = \avg { \dot{\alpha}_{ab}(0) \dot{\alpha}_{ab}(t) } \label{eqRaman5}
\end{equation}
and $a$ and $b$ indicate $x,y,z$.
We assess the accuracy of our IR results by comparing them to the experimentally measured band shifts and associated frequencies in the pure and co-crystalline phase~\cite{preston2012formation}. 
The IR spectra computed for the \aacrystal~co-crystal and the pure acetylene and ammonia crystals agree well with
the corresponding experimental spectra, Fig.~\ref{fig:IR}.
In addition to the absolute peak positions, the peak shifts that occur upon co-crystallization are
in reasonably good agreement with the experiment, Table~\ref{tab:IR}.
This good agreement between theory and experiment suggests that the density functional approximation used here can describe the \aacrystal~co-crystal with reasonable accuracy and yield accurate predictions for quantities like rotational correlation times. 
The blue shift of the \ace~$\nu_5$ asymmetric bend and the red shift of the $\nu_3$ asymmetric C-H stretch upon crystallization have been identified as characteristic signatures of co-crystal formation~\cite{preston2012formation},
and these shifts are captured by our predicted IR spectra.
The blue shifts of the co-crystal $\nu_5$ bands are slightly stronger in our simulation as compared to the observed experimental frequencies. 
Analogously, the red shift of the \ace~$\nu_3$ asymmetric stretch is more pronounced in our simulations than in experiments because of the large blue shift of the $\nu_3$ vibration in the pure phase with respect to the experimental $\nu_3$ frequency.  
Our computed spectrum also captures the relative shifts of the \amo~bands, however, given their small magnitude evident from the experimental shifts, they are rather subtle to observe.    
Notably, the blue shift in the \amo~$\nu_3$ vibrational band is accurately reproduced in our spectrum though the relative shift in frequency is overestimated.

\setlength{\tabcolsep}{8pt}
\begin{table}[tb]
\centering
 \begin{tabular}{c  c  c} 
 \hline  
 & \multicolumn{2}{c}{Frequencies (cm$^{-1}$)} \\ 
 \hhline{~|-|-} 
 &  \multicolumn{1}{c}{Calculated}  & \multicolumn{1}{c}{Experimental} \\ 
 \hhline{~-||-}
 Bands & Co-crystal & Co-crystal \\ [0.5ex] 
 \hline\hline
 \ace~$\nu_4$   &  702.3    &  669.2     \\  
                          &  748.2    &  721.9     \\
 \ace~$\nu_2$   &  1962.6  &  1944.4    \\
 \ace~$\nu_1$   &  3409.1  &  3300.8    \\
 \amo~$\nu_1$  &  3527.2  &  3368.8    \\
\hline
 \end{tabular}
 \caption{Calculated and experimental Raman vibrational frequencies of \ace~and \amo~in pure orthorhombic solid acetylene, cubic solid ammonia, and \aacrystal~co-crystal. Experimental Raman frequencies of pure and the co-crystalline phase are those reported by Cable \etal.~\cite{cable2018acetylene}}
\label{tab:Raman}
\end{table}%

\begin{figure}[tb]
\begin{center}
\includegraphics[width=0.48\textwidth]{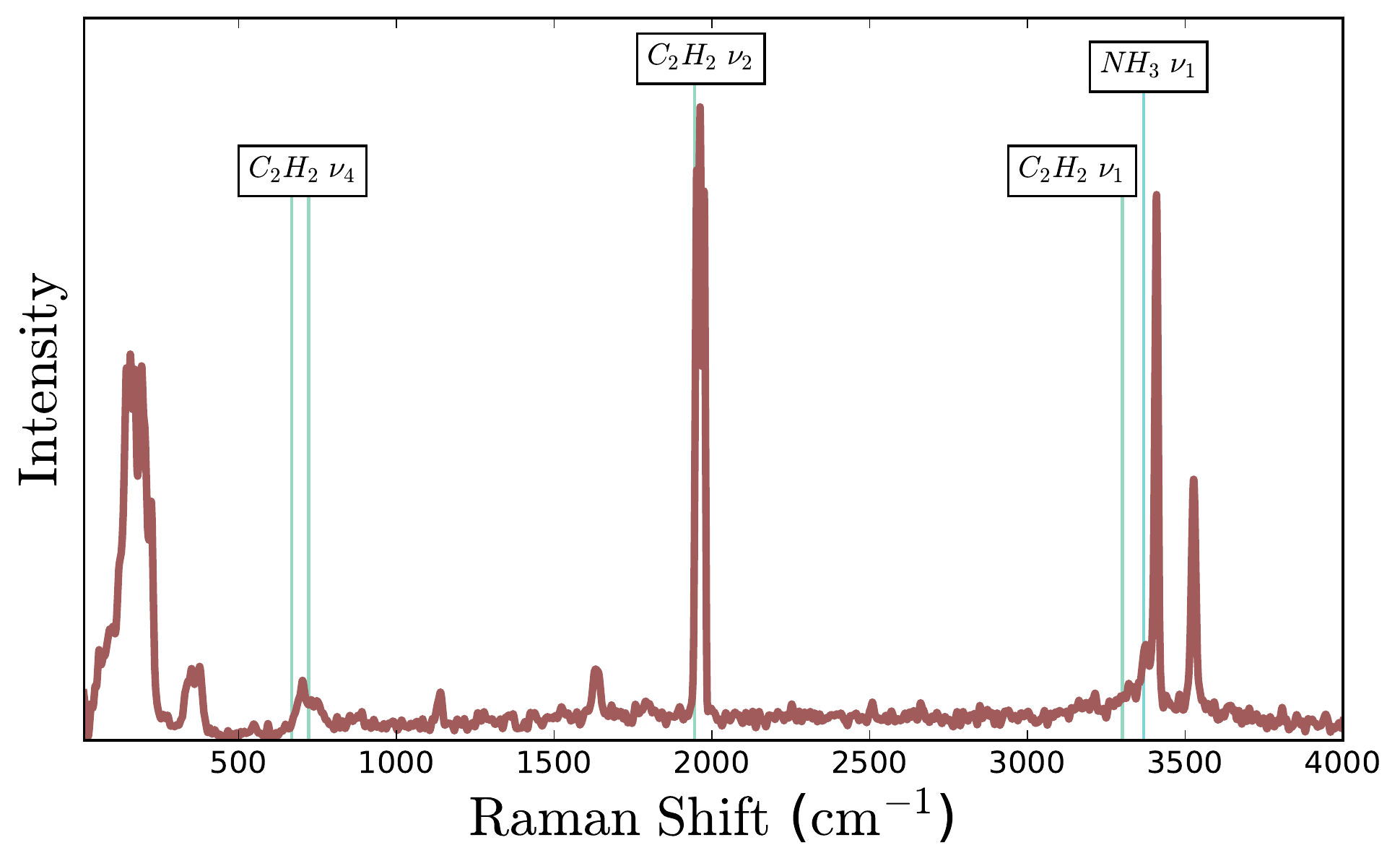}
\end{center}
\caption
{
Unpolarized Raman spectra computed for the \aacrystal~co-crystal at $T=90$~K.
Vertical lines indicate experimentally measured peak positions~\cite{cable2018acetylene}.
}
\label{fig:plotraman}
\end{figure}

%
In addition to IR, we also compute the Raman spectrum for the \aacrystal~co-crystal at 90~K, Fig.~\ref{fig:plotraman}. 
We compare the computed Raman signatures with the experimental bands reported by Cable~\etal~\cite{cable2018acetylene}.  
The peak positions predicted by our simulations are in good agreement with, but slightly higher frequency than those obtained experimentally,
Table~\ref{tab:Raman}. 
This level of agreement further validates the accuracy of our models and suggests that the range of predictions made regarding rotational timescales is reasonable.
%

\section*{Conclusions}

We have used \abint~molecular dynamics simulations to characterize the structure and dynamics of the \aacrystal~co-crystal. 
At 90~K, the temperature of Titan's surface, the \aacrystal~co-crystal is a plastic co-crystal.
Ammonia molecules within the co-crystal are orientationally disordered and undergo rotational motion on a picosecond timescale.
This rotational correlation time should be accessible in experiments through Raman and NMR spectroscopies~\cite{gordon1965molecular, gordon1965relations, rothschild1975dynamics}.
At 30~K, the \aacrystal~co-crystal is in a crystalline phase (translationally and orientationally ordered), such that a co-crystal--to--plastic co-crystal phase transition occurs between 30~and 90~K.
These conclusions agree well with the early experimental hints at the orientational disorder of ammonia molecules within the co-crystalline phase~\cite{boese2009synthesis}.  
The rotational dynamics of the ammonia molecules are dictated by the breakage and reformation of \nhpi~hydrogen bonds within the co-crystal, which we predict to have a lifetime on the order of a picosecond. 
The accuracy of our predictions is supported by the good agreement between experimentally measured and predicted IR and Raman spectra for the co-crystal.
We expect the \aacrystal~co-crystal to be one of many Titan cryominerals that exist in semi-disordered phases between crystal and liquid. 
Many of the multicomponent cryominerals that have been predicted are made of molecular components that exhibit orientationally disordered plastic crystal phases~\cite{maynard2018prospects,cable2021titan}, suggesting that these co-crystals might also exhibit these disordered phases. 
Understanding the existence of these disordered phases at Titan surface conditions is an essential step in understanding the geology of Titan from the molecular scale. 
For example, the dynamic disorder of plastic crystals can lead to significantly different mechanical properties than those of the ordered crystalline phase. 
In this context, the disorder of these phases must be accounted for when predicting mechanical properties, especially for large-scale modeling of surface geology.

We note that our ab initio simulations do not account for nuclear quantum effects. 
The quantum mechanical nature of nuclei, especially light nuclei, becomes increasingly important at low temperatures like those on the surface of Titan.
Future work will detail the importance of nuclear quantum effects in the molecular structure and dynamics of Titan's cryominerals. 
%

\begin{acknowledgements} 
This work is supported by the National Aeronautics and Space Administration
under grant number 80NSSC20K0609, issued through the NASA Exobiology Program.
We acknowledge the Office of Advanced Research Computing (OARC) at Rutgers,
The State University of New Jersey
for providing access to the Amarel cluster 
and associated research computing resources that have contributed to the results reported here.
This work used the Advanced Cyberinfrastructure Coordination Ecosystem: Services \& Support (ACCESS),
formerly Extreme Science and Engineering Discovery Environment (XSEDE)~\cite{towns2014xsede},
which is supported by National Science Foundation grant number ACI-1548562.
Specifically, this work used Stampede2 and Ranch at the Texas Advanced Computing Center through allocation TG-CHE210081.
\end{acknowledgements}

\bibliography{AAES}

\end{document}